\relax
\documentstyle [art12]{article}
\begin {document}
\bibliographystyle {plain}

\title{\bf The Toulouse Limit of the
Multi-Channel Kondo Model.}
\author {A. M. Tsvelik}
\maketitle
\begin {verse}
$Department~ of~ Physics,~ University~ of~ Oxford,~ 1~ Keble~ Road,$
\\
$Oxford,~OX1~ 3NP,~ UK$\\
cond-mat/9502081
\end{verse}
\begin{abstract}
\par
 We study the Toulouse limit of
the  multi-channel Kondo
model defined as the limit  of maximal anisotropy which can be
achieved without changing the
infrared behaviour of the  model. It is shown that when
the number of channels '$k$' exceeds two, the
interactions do not vanish and the Toulouse
limit is  described by a non-trivial field theory. Considerable
simplifications
take place only at $k = 2$ where the Bethe ansatz reproduces the  results
obtained by Emery and Kivelson [Phys. Rev. B {\bf 47}, 10812 (1992)].
\end{abstract}

PACS numbers: 72.10.Fk, 72.15.Qm, 73.20. Dx
\sloppy
\par
\section{Introduction}

 In his original paper Toulouse suggested that the anisotropic
spin-1/2 Kondo model
greatly simplifies at some special value of anisotropy, where it becomes a
model of free fermions$^1$. Later it was discovered that a similar point
exists  for the two-channel Kondo model, but the limiting theory is
a theory of free Majorana fermions$^{2,3}$. As we shall demonstrate below,
if the  number of channels '$k$' exceeds two, the Toulouse limit does
not correspond to  a theory  of free particles.
For example, it has been established by
Fabrizio and Gogolin$^4$ that for $k = 4$ the Toulouse limit
describes the boundary sine-Gordon model with the scaling
dimension of the boundary term $\Delta = 3/4$. The latter is a
non-trivial field theory whose infrared behaviour is governed by the
same scaling laws as for the isotropic 4-channel Kondo model$^{5}$.
Under  these circumstances one may wonder how to define the Toulouse
limit. We think that the logical choice is to define it
as the limit of maximal
anisotropy for which the infrared behaviour still remains the same as
for the isotropic model.
Since the Toulouse  limit plays a considerable role in
many
studies  of the Kondo models (see Refs. 2,3 and 6), we undertake to
study it in detail using the Bethe ansatz.

\section{Derivation of the Thermodynamic Equations}
 The Bethe ansatz equations for an
anisotropic k-channel Kondo model are similar to those for the
Heisenberg chain with spin $k/2$ and are given by (see, for example,
Refs. 6-8):
\begin{eqnarray}
[e_k(\eta; u_a)]^Ne_{2S}(\eta; u_a - 1/g) = \prod_{b = 1}^Me_2(\eta; u_a - u_b)
\label{eq}\\
 E = \sum_{a = 1}^M\frac{1}{2\mbox{i}}\ln e_k(\eta; u_a)\\
e_n(\eta; u) = \frac{\sinh[\eta(u - \mbox{i}n)]}{\sinh[\eta(u + \mbox{i}n)]}
\end{eqnarray}
where $S$ is the impurity spin, $g$ is the Kondo coupling constant,
$\eta$ is the anisotropy and N is the
length of the system and $M = kN/2 + S - S^z$ is the number of up
spins. The universal relationship beetween the
quantities $g$ and $\eta$ and the parameters of the Hamiltonian exists
only in the limit of weak anisotropy $\eta << 1$.
As we have mentioned above, the difficulty in determining
the Toulouse limit is resolved if we define it
as corresponding  to the maximal value of $\eta$ at which  the IR
fixed point of the
Kondo model still belongs to the same universality class as at $\eta
\rightarrow 0$. The periodicity of the trigonometric functions  in
Eqs.(\ref{eq}) suggests that for $k > 1$
the Toulouse limit is realized at $\eta =
\pi/2(k + 1/\nu)$ ($\nu \rightarrow \infty$). Here we percieve a
profound difference between the single-channel ($k$ = 1) and
multi-channel Kondo models. In the single-channel case the Toulouse
limit is defined as the point where the scattering vanishes. It occurs at
$\eta
= \pi/4$ where the right hand side of Eqs.(\ref{eq}) becomes 1.
At $k > 2$ the point $\eta = \pi/4$ cannot
be continuously connected with the isotropic point $\eta \rightarrow
0$, since at $\eta =
\pi/2k$ the left hand side becomes equal to 1, which marks a
discontinuity. Therefore the Toulouse limit defined as the  limit of
maximal anisotropy is no longer equivalent to  free particles and
corresponds to a non-trivial field theory. Thus, as we have already
mentioned,  according to the
arguments given in Ref. 4, at $k = 4$ the Toulouse limit
describes the boundary sine-Gordon model with the scaling dimension of
the boundary term $\Delta = 3/4$.

 The Toulouse limit has been
considered in Refs. 7-9  in the context of the model of the
Anisotropic Principal
Chiral Field. The most  detailed derivation  of the thermodynamic
Bethe ansatz (TBA)
equations is given in Ref. 9. The equations for the Kondo model are
essentially the same as for the Principal Chiral Field, but,
nevertheless, some details need explanation which is presented in this
article. We will show that  the TBA equations in the
Toulouse limit have the following form:
\begin{eqnarray}
\epsilon_n = s*\ln\left(1 + \mbox{e}^{\epsilon_{n - 1}}\right)\left(1 +
\mbox{e}^{\epsilon_{n + 1}}\right) \nonumber\\
+
\delta_{n,k - 2}s*\ln
\left(1 + \mbox{e}^{\epsilon_{k}}\right) + \delta_{n,k - 1}s*f(v), \: n = 1,
... k - 1\\
\epsilon_{k} = s*\ln\left(1 + \mbox{e}^{\epsilon_{k - 2}}\right)
- 2\exp( - \pi v/2) - s*f(v)\label{eq1}\\
F_{imp} = - T\int_{-\infty}^{\infty} \mbox{d}v s(v + \frac{2}{\pi}\ln
T_K/T)\ln\left[1 +
\mbox{e}^{\epsilon_{2S}(v)}\right]\label{eq3}
\end{eqnarray}
where
\[
s*g(v) = \int_{-\infty}^{\infty}du\frac{g(u)}{4\cosh[\pi(v - u)/2]},
\]
the function $f(v)$ is defined later in the text (see Eq.(\ref{great})
and the Appendix)
and the Kondo temperature $T_K$ is defined as:
\begin{equation}
T_K = \lim_{\nu, \Lambda \rightarrow \infty}\frac{\pi\Lambda}{2\nu}\exp(-
\pi/2g)\label{tk}
\end{equation}
At zero magnetic field $f(v) = 0$.
When taking  the Toulouse limit one has to be careful to keep
the Kondo energy scale,
$T_K$, finite. This is because at $\nu \rightarrow \infty$
the left hand side of Eqs.(\ref{eq}) formally becomes equal to one. In
order to get non-trivial solutions one has to renormalize
the cut-off, $\Lambda \sim \nu$.

 As it was shown by Takahashi and Suzuki$^{10}$ for $k = 1$,
and later for general $k$ by Kirillov and Reshetikhin$^8$,
at  $\eta = \pi/(k + 1/\nu)$ Eq.(\ref{eq}) has  the following string
solutions: (i) there are $k - 1$ strings of lengths $n = 1, 2, ... k -
1$ whose energies we denote as $\epsilon_n$, (ii) there are kink and
antikink excitations $E_{\sigma}, \: \sigma = \pm$, (iii) there are
$\nu - 1$ breathers with energies $\kappa_j, \: j = 1, 2, ... \nu -
1$. The total number of excitation branches is equal to $k +
\nu$. Kinks and antikinks have spin 1/2 with a Landee  factor which is
a function of the interactions, whilst the breathers are
spinless.

 The TBA equations for the strings are:
\begin{eqnarray}
\epsilon_n = s*\ln\left(1 + \mbox{e}^{\epsilon_{n - 1}}\right)\left(1 +
\mbox{e}^{\epsilon_{n + 1}}\right)\nonumber\\
 +
\delta_{n,k - 1}\left[s*\sum_{\sigma}\ln
\left(1 + \mbox{e}^{- E_{\sigma}}\right) + \sum_j\xi_j*\ln\left(1 +
\mbox{e}^{\kappa_j}\right)\right], \: n = 1, ... k - 1
\end{eqnarray}
where
\[
\xi_j(\omega) = \frac{\cosh[(1 - j/\nu)\omega]}{\cosh \omega}
\]
In the limit $\nu \rightarrow \infty$ this kernel becomes a
delta-function.

 The TBA equations for the breathers are:
\begin{eqnarray}
\ln\left(1 + \mbox{e}^{\kappa_j}\right) - B_{jm}*\ln\left(1 + \mbox{e}^{-
\kappa_m}\right) = \nonumber\\
\frac{2\nu}{\pi}\sin(\pi j/2\nu)\mbox{e}^{- \pi v/2} + b_j*\sum_{\sigma}\ln
\left(1 + \mbox{e}^{- E_{\sigma}}\right) - \xi_j*\ln\left(1 +
\mbox{e}^{\epsilon_{k - 1}}\right) \label{bre}
\end{eqnarray}
where the Fourier transforms of the kernels are given by:
\begin{eqnarray}
B_{jm}(\omega) = 2\coth(\omega/\nu)\frac{\cosh[(1 -
j/\nu)\omega]\sinh(m\omega/\nu)}{\cosh \omega}, \: (j > m)\nonumber\\
B_{jm} = B_{mj}\nonumber\\
b_j(\omega) = \frac{\coth(\omega/\nu)\sinh(j\omega/\nu)}{\cosh\omega}
\end{eqnarray}
In the limit $\nu \rightarrow \infty$ we have
\begin{equation}
B_{jm}(v) = 2\min(j,m)\delta(v), \: b_j(v) = 2j s(v)
\end{equation}
In this limit Eqs.(\ref{bre}) can be solved and the functions $\xi_j$
expressed in terms of other excitation energies.

The TBA equations for kink and anti-kink excitations are
\begin{eqnarray}
E_{\sigma}(v) - \sum_{\sigma'}K*\left(1 + \mbox{e}^{-
E_{\sigma}}\right) = \nonumber\\
\frac{\nu}{\pi}\mbox{e}^{- \pi v/2} - g_L\sigma H/T - \sum_js*\ln\left(1 +
\mbox{e}^{
\epsilon_j}\right) + \sum_jb_j*\ln\left(1 + \mbox{e}^{-
\kappa_j}\right) \label{kink}
\end{eqnarray}
where
\[
K(\omega) = \frac{\sinh[(1 -
\nu^{-1})\omega]}{2\cosh\omega\sinh(\omega/\nu)}
\]

 The free energy has the following form:
\begin{eqnarray}
\frac{1}{L}F_{bulk} = \nonumber\\
-
\frac{\nu}{\Lambda} T^2\int_{-\infty}^{\infty}
\mbox{d}v \mbox{e}^{- \pi v/2}\left[\frac{1}{\pi}\sum_{\sigma}\left(1 +
\mbox{e}^{-
E_{\sigma}}\right) + \nu^{-1}\sum_j  j\ln\left(1 + \mbox{e}^{-
\kappa_j}\right)\right]\label{bfree}\\
F_{imp} = - T\int_{-\infty}^{\infty}\mbox{d}v s[v +
\frac{2}{\pi}\ln(T_K/T)]\ln\left(1 + \mbox{e}^{
\epsilon_{2S}(v)}\right)
\end{eqnarray}
where $\Lambda = N/L$ and $L$ is the length of the system. The Kondo
temperature $T_K$ is defined by Eq.(\ref{tk}).

 Now we shall consider the limit $\nu \rightarrow
\infty$. As we have said above,
we shall keep $T_K$ finite  when taking  this limit, which  implies
keeping the combination $\Lambda/\nu$ finite.
Let us consider the equations for $E_{\sigma}$ first. At  $\nu \rightarrow
\infty$ the kernel $K(\omega) \rightarrow \nu \tilde K(\omega)$ where
\[
\tilde K(\omega) = \frac{\tanh\omega}{2\omega}
\]
Since the right hand side of Eq.(\ref{kink}) is proportional to
$\nu$, $E_-(v)$ goes to infinity and
only $E_+(v)$ remains finite and only in a finite
magnetic field. The corresponding equation is
\begin{equation}
\tilde K*\ln\left(1 + \mbox{e}^{- E_+(v)}\right) = \tilde g_LH/2T -
\frac{1}{\pi}\mbox{e}^{- \pi v/2} \label{eq:K}
\end{equation}
where
\begin{equation}
\tilde g_L = \lim_{\nu \rightarrow \infty}\frac{g_L}{\nu}
\end{equation}
The latter quantity is the effective impurity
Landee factor which is  difficult to determine in the Bethe
ansatz framework. Similar  difficulties are  common for all models with
linear spectrum and originate from the fact that in the Bethe ansatz
one chooses  as the reference state
a ferromagnetic state with a very high energy
(see the discussions in Ref. 11). According to  Refs. 3 and 6 the
impurity magnetic susceptibility vanishes in the Toulouse limit, which
indicates that $\tilde g_L = 0$. However, this result was obtained
under the condition that the impurity and the conduction electrons
have the same Landee factors. For the general
case  $g_{imp} \neq g_c$, where $g_{imp}$ and $g_c$ are
the bare Landee factors of the impurity
and the conduction electrons,  it was demonstrated
by Fabrizio {\it et al.}$^{12}$ that
\begin{equation}
\tilde g_L = g_{imp} - g_c
\end{equation}
It make sense to consider $g_{imp} \neq g_c$ and keep $\tilde g_L$ finite.
In what follows we shall denote $f(v) = \ln\left(1 + \mbox{e}^{-
E_+(v)}\right)$. The solution of Eq.(\ref{eq:K}) is given in the Appendix.

 The $\nu \rightarrow \infty$ limit of the TBA equations for breathers are
given by
\begin{eqnarray}
\ln\left(1 + \mbox{e}^{\kappa_j(v)}\right) - 2\sum_m\min(j,m)\ln\left(1 +
\mbox{e}^{-
\kappa_m(v)}\right) = \nonumber\\
j\left(\mbox{e}^{- \pi v/2} + 2s*f(v)\right) - \ln\left(1 +
\mbox{e}^{\epsilon_{k - 1}(v)}\right) \label{limbr}
\end{eqnarray}
Inverting the kernel we transform these equations to the more
familiar form:
\begin{eqnarray}
\kappa_j(v) = \frac{1}{2}\ln\left(1 + \mbox{e}^{\kappa_{j -
1}(v)}\right)\left(1 + \mbox{e}^{\kappa_{j + 1}(v)}\right) -
\frac{1}{2}\delta_{j,1}\ln\left(1 +
\mbox{e}^{\epsilon_{k - 1}(v)}\right)\nonumber\\
\lim_{j \rightarrow \infty}\frac{\kappa_j(v)}{j} = \left(\mbox{e}^{-
\pi v/2} + 2s*f(v)\right) \equiv \epsilon_0(v)
\end{eqnarray}
The general solution of this  system was obtained by Takahashi$^{13}$:
\begin{eqnarray}
1 + \mbox{e}^{\kappa_{j}} = \left\{\frac{\sinh[\frac{1}{2}\epsilon_0(j
+ 1) - u(v)]}{\sinh(\frac{1}{2}\epsilon_0)}\right\}^2
\end{eqnarray}
In our case the function $u(v)$ is determined by the boundary
condition at $j = 1$:
\begin{equation}
1 + \mbox{e}^{\epsilon_{k - 1}} =
\left[\frac{\sinh(\frac{1}{2}\epsilon_0)}{\sinh(\frac{1}{2}\epsilon_0 -
u)}\right]^2
\end{equation}
Using this result we can calculate the infinite sums appearing in
other TBA equations:
\begin{eqnarray}
\sum_j\ln\left(1 + \mbox{e}^{\kappa_{j}}\right) =
\ln\left[\frac{\mbox{e}^{-\epsilon_0/2}\sinh(\epsilon_0 -
u)}{\sinh(\frac{1}{2}\epsilon_0 - u)}\right]\label{equ}\\
\sum_jj\ln\left(1 + \mbox{e}^{\kappa_{j}}\right) =
- \ln\left(1 - \mbox{e}^{2u -\epsilon_0}\right) \nonumber\\
\equiv \ln\left(1 + \mbox{e}^{\epsilon_k}\right)/\left(1 -
\mbox{e}^{\epsilon_0}\right)\label{cont}
\end{eqnarray}
Here we introduced the new energy $\epsilon_k$ in order to get rid of
$u$. We shall also redefine $\epsilon_{k - 1}$ introducing a new
function $\epsilon_{k - 1}'$:
\begin{equation}
1 + \mbox{e}^{\epsilon_{k-1}} = \left(1 +
\mbox{e}^{\epsilon_k}\right)\left(1 + \mbox{e}^{\epsilon_{k -
1}'}\right)\label{eps}
\end{equation}
Substituting Eqs.(\ref{equ}),(\ref{cont}) and Eq.(\ref{eps}) into
Eqs.(9) we obtain Eqs.(\ref{eq1}),  and (\ref{eq3}) with
$f(v)$ given by Eq.(\ref{eq:K}).

\section{Thermodynamic Properties in Zero Magnetic Field}

 The system of
TBA equations (\ref{eq1}) and (\ref{eq3}) really have
different topologies at $k = 2$ and $k > 2$. In the former case the
equations decouple  and we have
\begin{eqnarray}
\epsilon_2(v) = - 2\exp(- \pi v/2) - s*f(v)\nonumber\\
\epsilon_1 = s*f(v)
\end{eqnarray}
The function $f$ is independent of $\epsilon_{1,2}$, the explicit
expression for it is derived in the Appendix (see Eq.(38)). Substituting
Eq.(38) into Eq.(6) we obtain the following expression for  the impurity free
energy:
\begin{eqnarray}
F_{imp} = - T\int \mbox{d}vs[v + \frac{2}{\pi}\ln T_K]\ln\left\{1 +
\exp[ s*f(v + \frac{2}{\pi}\ln T)]\right\}\nonumber\\
 = - \frac{T}{\pi}\int_0^{\infty} \frac{\mbox{d}x}{1 + x^2}\ln\left\{1 +
\exp[\frac{T_K}{2T}(\sqrt{(\tilde g_LH/T_K)^2 + x^2} - x)]\right\}
\label{impur}
\end{eqnarray}
This formula describes the free energy of a Majorana fermion with
dispersion $E(v)$. Since $f \sim H$,  $E(v)$ vanishes in zero magnetic
field together
with the impurity part of the specific heat. Thus we reproduce the results
obtained in Refs. 2,3 and 6.

 At $k > 2$ no simplifications take place. Neither the impurity
contribution to the specific heat disappears at $H = 0$, and
the Toulouse limit is described by a non-trivial field theory.
We can solve the TBA equations
only approximately for   $T >> T_K$ and $T<< T_K$.
At $T << T_K$ we can expand  Eqs. (\ref{eq1}) and
(\ref{eq3}) in powers of $g_k \equiv \ln\left[1 +
\exp(\epsilon_k)\right]$ and $f$, as was done in Ref. 14. The answer
is:
\begin{eqnarray}
s*\left(1 + \mbox{e}^{\epsilon_{2S}(v - \frac{2}{\pi}\ln(T_K/T) )}\right)
= \frac{\Phi(k - 1)}{\Phi(k - 2)\Phi(2S)}\nonumber\\
\times[\Phi(2S - 1)\hat l_{2S + 2}
- \Phi(2S + 1)\hat l_{2S}]*\left[g_k(v) + \frac{\Phi(k -
2)}{4\cos^2(\pi/k + 2)}s*f(v)\right]\nonumber\\
l_n(\omega) = \frac{\sinh n\omega}{\sinh(k + 2)\omega}
\end{eqnarray}
where
\[
\Phi(n) = \frac{\sin\left[\pi (n + 1)/(k + 2)\right]}{\sin\left[\pi /(k +
2)\right]}
\]

The first non-vanishing pole of the kernel is at $\omega =
-2\pi\mbox{i}/(k+ 2)$ which  gives the correct low temperature
asymptotics for the impurity free energy:
\begin{eqnarray}
F_{imp} = - AT(T/T_K)^{4/(k + 2)}\nonumber\\
A = \int \mbox{d}v \mbox{e}^{- 2\pi v/(k + 2)}[g_k(v) + \frac{\Phi(k -
2)}{4\cos^2(\pi/k + 2)}s*f(v)]
\end{eqnarray}

\section{Thermodynamic Properties at T = 0}

 \begin{equation}
f^{(-)}(v) + \int_Q^{\infty}K(v - u)f^{(+)}(u) \mbox{d}u =
\frac{\tilde g_LH}{2T} - \frac{1}{\pi}\mbox{e}^{- \pi v/2} \label{great}
\end{equation}
The solution of this equation is obtained in the Appendix.
Substituting Eq.(\ref{f}) into the expression for the impurity free
energy (\ref{impur}) taken at $T \rightarrow 0$
and differentiating the result with respect to the magnetic field,
we get the following expression for the impurity magnetic moment:

\begin{equation}
M_{imp} = \frac{k^2}{\pi}\int\frac{\mbox{d}x \sin(2S\pi/k)}{\cosh\pi
x -  \cos(2S\pi/k)}\frac{\tilde g_LH}{\sqrt{{\tilde g_L}^2H^2 + T_K^2\exp(- \pi
kx)}}
\end{equation}
For $k = 2, \: S = 1/2$ this integral can be calculated:
\begin{eqnarray}
M_{imp} = \nonumber\\
= \frac{h}{\pi\sqrt{|1 - h^2|}}\left[
\ln\left(h^{-1} + \sqrt{h^{-2} - 1}\right)\theta(1 - h) +
\sin^{-1}\sqrt{1 - h^{-2}}\theta(h - 1)\right]
\end{eqnarray}
 For $k = 4, \: S = 1$ the integral for the magnetization also looks
rather simple, but, nevertheless, cannot be expressed in elementary functions:
\begin{equation}
M_{imp} = 4\sqrt h\int_0^{\infty} \frac{\mbox{d}y}{\sqrt{y(1 + y^2)}(1
+ hy)}
\end{equation}
In both cases $h = \tilde g_LH/T_K$.

\section{Conclusion}

 Here we just give a brief summary of the picture described in the
main text. It turns out that the Toulouse limit is very different for
$k = 2$ and $k > 2$. In the former case one can solve the TBA
equations explicitly for any temperatures and magnetic fields. It is
also true that at $H = 0$ the impurity contribution to the free energy
becomes trivial $F_{imp} = - T\ln\sqrt 2$. This result holds for any
$H$ when  $g_{imp} = g_c$ (see Refs. 3, 6). Nothing of that sort
happens at $k > 2$. The TBA equations remain non-trivial and no
explicit analytic solution is available.

\section{Acknowledgements}
 The author expresses his gratitude to A. Gogolin for very helpful
critical remarks and interest to the work and to P. de Sa for his help in
preparation of the manuscript.

\section{Appendix}

 Eq.(\ref{eq:K}) is quite special in that respect that its solution can be
expressed in elementary functions. Namely, one can treat this equation
as a limiting case of the equation
\begin{equation}
\int_{-Q}^Q\mbox{d}u\ln|\coth[\frac{\pi}{4}(v - u)]|\tilde f(u) = - \pi
\tilde g_LH
+ 4m\cosh(\pi v/2)
\end{equation}
The latter  equation becomes Eq.(\ref{eq:K}) at
$m \rightarrow 0, \: m\exp(\pi Q/2) = $const with
\[
\tilde f(v) = Tf(v +
\frac{2}{\pi}\ln T)
\]
Differentiating this equation with respect to $v$ and using the
substitution
\[
\tilde f(u) = [\cosh(\pi u/2)]^{-1}F(u), \: x = \tanh(\pi u/2)
\]
we reduce it to the canonical form
\begin{equation}
P\int_{-B}^B\frac{\mbox{d}y F(y)}{x - y} = g(x);\: \: g(x) =
- \pi m\frac{x}{1 - x^2}
\end{equation}
This equation has the following general solution$^{15}$:
\begin{equation}
F(x) = - \frac{1}{\pi^2}\sqrt{B^2 - x^2}P\int_{-B}^B\frac{g(y)\mbox{d}y
}{(y - x)\sqrt{B^2 - y^2}}
\end{equation}
In our case we get the following expression for $\tilde f(u)$:
\begin{equation}
\tilde f(u) = - m\frac{\left[B^2\cosh^2(\pi u/2) - \sinh^2(\pi
u/2)\right]^{1/2}}{(1 - B^2)^{1/2}} \label{sol}
\end{equation}
The limit $B$ can be  determined from the original equation. In
particular, substituting there the solution (\ref{sol}) and putting
$v = 0$ we obtain the following condition:
\begin{eqnarray}
\int_{-Q}^Q\mbox{d}v\ln|\coth (\pi v/4)|\left[\sinh^2(\pi
Q/2)\cosh^2(\pi v/2) - \cosh^2(\pi
Q/2)\sinh^2(\pi v/2)\right]^{1/2} \nonumber\\
= - 4 + \pi H\tilde g_L/m \label{con}
\end{eqnarray}
where $B = \tanh(\pi Q/2)$. In the limit $H/m \rightarrow \infty$ when
$Q \rightarrow \infty$ we
obtain from Eq.(\ref{con}):
\[
m\mbox{e}^{Q\pi/2} = \tilde g_LH
\]
and
\begin{eqnarray}
s*\tilde f = \frac{\tilde g_LH}{2}\left(- \mbox{e}^{- \pi v/2}
+ \sqrt{1 + \mbox{e}^{- \pi v}}\right) \label{f}
\end{eqnarray}

\end{document}